\documentclass[aip,pop,reprint,showpacs,notitlepage,twocolumn]{revtex4-1}
\usepackage{amsmath}
\usepackage{bm}
\usepackage{amssymb}
\usepackage{graphicx}
\usepackage[english]{babel}
\usepackage{pdfsync}
\usepackage{mathptmx}
\usepackage{helvet}

\begin{document}

\title{From extended phase space dynamics to fluid theory}

\author{Jens Zamanian, Martin Stefan, Mattias Marklund, Gert Brodin} 

\affiliation{Department of Physics, Ume{\aa} University, SE--901 87 Ume{\aa}, Sweden}

\begin{abstract}
We derive a fluid theory for spin-1/2 particles starting from an extended kinetic
model based on a spin-projected density matrix formalism. The evolution equation for the spin density is found to
contain a pressure-like term. We give an example where this term is important by looking at a linear
mode previously found in a spin kinetic model.  
\end{abstract}

\maketitle

\section{Introduction}

Quantum plasma physics is an active field of research and the methods have been applied to investigate a wide range of different effects. Some recent examples are quantum ion-accoustic waves \cite{bains}, Jeans instabilities in quantum plasmas \cite{ren}, trapping effects \cite{shah}, magnetization by photons \cite{shukla} and relativistic effects \cite{melrose,masood}. Typically, the quantum hydrodynamic equations are derived starting from the Schr\"odinger equation and making the Madelung ansatz\cite{manfredi-howto}. However, a method that more closely resembles the classical case is to use kinetic equations as a starting point, see Ref.\ [\onlinecite{manfredi-howto}] for a comparison between the different approaches. 
The field of quantum kinetic theory\cite{carruthers} in many ways started with the ambitions of Wigner, as presented in Ref.\ [\onlinecite{wigner32}], to bridge the gap between classical Liouville theory and statistical quantum dynamics \cite{weyl,groenewold,moyal}. Thus, the development of quantum kinetic theory was partly due to an interest in obtaining a better understanding of the quantum-to-classical transition \cite{zachos}. However, another important aspect of quantum kinetic theory is as a computational tool for, e.g., quantum plasmas \cite{shukla-eliasson}, condensed matter systems \cite{Markowich,haug-jauho}, and, in general, quantum systems out of equilibrium \cite{rammer}, and in that respect shares many commonalities with quantum optics \cite{leonhardt}.

Systems where the electron spin plays an important role is a growing field of research. One reason for this is the promise of novel electronic devices. Furthermore, the utilization of spin may also play a key role in the yet-to-come 
implementation of quantum computing \cite{wolf2001,zutic2004}. There is also a growing interest in quantum effects in laser generated plasmas \cite{marklund-lundin2009,eliezer,glenzer} and other high energy density environments \cite{drake}. Thus, for such system, more encompassing modeling is required as compared to the condensed matter system used in modeling of, e.g., electronic devices. 
Spin effects has also been studied in the context of thermonuclear fusion plasmas by Cowley, Kulsrud and Valeo \cite{cowley}, who investigated spin depolarization rates. Their idea was to start from a kinetic theory for spin-1/2 particles and calculate moments of this to obtain macroscopic quantities. The distribution function in their theory was defined on an extended phase-space where the spin degree of freedom was described by vector. The kinetic equation, however, was not derived from a fully quantum mechanical theory and this is a point that later was criticized by Zhang and Balescu\cite{zhang}, who derived a fully quantum mechanical equation for spin-1/2 particles. Zhang and Balescu also argued that the spin cannot be described by an extended phase-space and thus their kinetic approach is based on a Wigner matrix. However, it is possible to describe the spin through an extended phase-space, see \cite{scully} for a review of quasi-distribution function for spin. In an subsequent paper\cite{balescu} Balescu and Zhang then went on to derive a fluid moment theory based on their kinetic theory. This theory was derived to leading order in $\hbar$ which means that the dynamics due to the spin dipole force is not retained. 

Recently, we have derived a scalar kinetic theory for spin-1/2 particles \cite{brodin2009,zamanian2010}, where the spin is described by a unit vector in the distribution function. This is natural extension of regular quantum kinetic theories \cite{kremp-etal2005} as well as fluid models with spin (see e.g.\ [\onlinecite{marklund2007}] and [\onlinecite{brodin-marklund2007}]). This theory has the advantage of being of close mathematical resemblance to the Vlasov theory for a classical system, where the spin degree of freedom is seen to correspond to a classical magnetic dipole moment. This may guide the intuition when interpreting the corresponding fluid moment theory. It may also make classical numerical codes amendable and adaptable for quantum kinetic simulations. The scalar quantum kinetic theory \cite{zamanian2010} is seen correspond to the kinetic equation of Cowley, Kulsrud and Valeo (in the collisionless limit) to leading order in $\hbar$. Our starting point in this paper is a kinetic equation valid to order $\hbar^2$, which will then keep not only precession but also the dipole force acting on the spin. 

Spin fluid models have also been applied in other contexts, for example considering ultra cold trapped atomic gases \cite{du2008}. In this case, the atoms interact via two hyperfine states and these can be viewed as the two spin states, hence the important feature of such a theory
is the spin-spin interaction. In the theory considered here, however, we will not deal with 
two-particle spin-spin interactions, but rather use a mean-field approach. 

In the presence of a vector potential, the gauge issue makes itself reminded, since a gauge transformation $A_\mu(x) \rightarrow A_\mu(x) + \partial_\mu\chi(x)$ of the four-vector potential gives the transformation $\psi(x) \rightarrow \psi(x)\exp[ie\chi(x)/\hbar c]$ of the wave function. Thus, in order to obtain a gauge invariant quasi-distribution, the ansatz for the Wigner function has to be modified. There is infinitely many ways to do this and external criteria have to be applied in order to choose a specific modification \cite{dirac1955}. Thus, there are different `standard routes' for the modification due to the vector potential, e.g. using a Wilson loop \cite{lavelle-mcmullan1997}. In quantum kinetic theory the standard way is, to the best of our knowledge, best represented by Stratonovich and others \cite{stratonovich1956,serimaa-etal1986}, who choose to take the gauge modification of the wave function into account by adding a line integral of the vector potential to the canonical momenta in the definition of the Wigner function. However, the corrections due to the gauge choice enters the phase space dynamics at higher order in the Planck's constant $\hbar$. As we in this paper will be mainly interested in the lowest order spin corrections to the fluid equations, we intend not to make use of the full gauge operators that comes from short length scale (compared to the characteristic de Broglie wavelenght) contributions. The fluid equations based on the full kinetic theory has been investigated recently in the spinless case\cite{haas2010-1,haas2010-2}.

In Section 2 we derive the fluid moments of the spin kinetic equation and discuss the derived model. The equations are then coupled via Maxwell's equations and we consider linear modes of the system in Section 3. We also briefly discuss the closure problem. In Section 4 we draw our conclusions and discuss some future applications.

\section{Moments of the spin kinetic model} 
In a recent paper \cite{zamanian2010} we derived a
kinetic model for a particle with spin-1/2. In this theory the system is described by a
quasi-distribution function $f(\mathbf x, \mathbf v, \mathbf s,t)$ which is a combination of the
Wigner function \cite{wigner32} in phase-space variables $\mathbf x$ and $\mathbf v$ and a
transformation for the spin variable, the unit vector $\mathbf s$. It has the property that the marginal distribution $n(\mathbf x, \mathbf s) = \int d^3 v f$ yields the probability density to find a particle in position $\mathbf x$ with spin in the direction given by $\mathbf s$. Similarly $n(\mathbf v, \mathbf s) = \int d^3 x f$ gives the probability to find a particle with a velocity $\mathbf v$ and spin in the $\mathbf s$-direction.
A third important property is that the magnetization $\mathbf M$ due to the spin is obtained by
\begin{equation}
	\mathbf M(\mathbf x, t) = 3 \mu \int d\Omega \mathbf s f(\mathbf x, \mathbf v, \mathbf s, t), 
\end{equation}
where $d\Omega = d^3 v d^2 s$ and $\mu$ is the magnetic moment of the particles. Here we will
consider electrons and hence $\mu = - (g/2) e\hbar / (2m)$ where $e$ is the elementary charge, $m$
is the electron mass and $\hbar$ is Planck's constant divided by $2\pi$. The g-factor for an
electron is modified from the value $g=2$, predicted by the Dirac theory, to $g \approx 2.0023$ when
one takes vacuum fluctuations into account, see Ref.\ \cite{odom2006} for a recent measurement.  
Note the factor 3 which is related to the non-commutation of the different spin components. 

If we consider scale lengths longer than the characteristic de Broglie wavelength the evolution equation for the
quasi-distribution function is found to be \cite{zamanian2010} 
\begin{eqnarray}
	&&\partial_t f + \mathbf v \cdot \nabla_x f + 
	\frac{q}{m} (\mathbf E + \mathbf v \times \mathbf B ) \cdot \nabla_v f
\nonumber \\&& 
	+ \frac{\mu}{m} \nabla_x 
	\left[ \left( \mathbf s + \nabla_{s} \right) \cdot \mathbf B \right] \cdot \nabla_v f
	+ \frac{2 \mu}{\hbar} (\mathbf s \times \mathbf B) \cdot \nabla_s f
	= 0.
	\nonumber \\
	\label{vlasov}
\end{eqnarray}
The terms proportional to $\mu$ are terms that arises due to the spin, the first term is the dipole force for a
particle with magnetic moment $\mu \hat{\mathbf{s}}$, which has been modified due to the quantum
mechanical nature of the spin, by the term $(\mu/m) \nabla_x(\mathbf B \cdot \nabla_s) \cdot
\nabla_v f$. The last term on the second row is due to spin precession.   
To obtain a macroscopic theory for the system we define the fluid moments
\begin{eqnarray}
	n &=&  \int d\Omega f ,
	\label{density} \\ 
	U_i &=& \frac{ 1}{n} \int d\Omega v_i f ,
	\label{velocity} \\
	P_{ij} &=& m \int d\Omega (v_i - U_i) (v_j - U_j) f ,
	\label{pressure} \\
	S_i &=&  \frac{ 3}{n}   \int d\Omega  \hat s_i f ,
	\label{spindens} \\ 
	\Sigma_{ij} &=& m \int d\Omega (3 \hat s_i - S_i) (v_j - U_j) f ,
	\label{spinvel} \\
	\Lambda_{ijk} &=& m \int d\Omega (3 s_i - S_i) (v_j - U_j) (v_k - U_k) f .
	\label{svv}  
\end{eqnarray}
Note that these definitions are in line with the results of Ref.\ [\onlinecite{balescu}]. However, our definition of the mixed spin-velocity moment \eqref{spinvel} contains only the thermal contribution.
The first three moments, Eqs.\ \eqref{density} to \eqref{pressure}, are the density, velocity and
pressure. The fourth moment, Eq.\ \eqref{spindens}, yields the spin density per volume. The two
following moments are mixed spin and velocity moments. They have no straightforward physical  interpretation, but we will see that the spin-velocity moment \eqref{spinvel} will act as a pressure-like term in the evolution equation for the spin density. 

Using the definition for the different moments above it is straightforward to derive evolution equations using Eq.\ \eqref{vlasov}. The evolution of the density moment is given by the continuity equation 
\begin{equation}
	\partial_t n + \nabla_x \cdot ( n \mathbf U) = 0 .
	\label{9}
\end{equation}
The fluid velocity is evolving according to  
\begin{eqnarray}
        \frac{D U_i}{Dt} =     
	\frac{q}{m} ( E_i + \epsilon_{ijk}  U_j B_k )
	+ \frac{\mu}{m} S_j \frac{\partial B_j}{\partial x_i}   
	- \frac{1}{n m} \frac{\partial P_{ij}}{\partial x_j}  ,  
	\label{10}
\end{eqnarray}  
where summation over repeated indices is implied and the convective derivative $D/Dt \equiv \partial/\partial t + U_j
\partial/\partial x_j$ has been introduced. 
The second term on the right hand side is a magnetic dipole force due to the spin. 
The pressure moment is evolving according to 
\begin{eqnarray}
  \frac{ D P_{ij}}{Dt} &=&  
  - P_{ik} \frac{ \partial U_j}{\partial x_k}  
  - P_{jk} \frac{\partial U_i}{\partial x_k}    
  - P_{ij} \frac{\partial U_k}{\partial x_k} 
  \nonumber \\ && 
  + \frac{ q}{m} \epsilon_{imn} P_{jm} B_n 
  + \frac{ q}{m} \epsilon_{jmn} P_{im} B_{n}
  \nonumber \\ && 
  + \frac{\mu}{m} \Sigma_{ik} \frac{\partial B_k}{\partial x_j}  
  + \frac{ \mu }{m} \Sigma_{jk} \frac{\partial B_k}{\partial x_i}  
  - \frac{\partial Q_{ijk}}{\partial x_k} 
\label{pressev}
\end{eqnarray}
and as expected has some extra terms due to the spin (cf.\ [\onlinecite{haas2010-1}]). 
The evolution equation for the spin density is given by 
\begin{eqnarray} \label{12}
  \frac{ D S_i}{Dt} =  
	\frac{2\mu}{\hbar} \epsilon_{ijk} S_j B_k  
	- \frac{1}{nm} \frac{\partial \Sigma_{ij}}{\partial x_j} . 
\end{eqnarray}
Comparing this with Eq.\ \eqref{10} we see that this mixed spin-velocity tensor is acting as a 
pressure term on the spin moment. Exactly which physics is captured by this term is mainly a subject 
of further research. In this paper we give an example where this term is important. 
Finally, the spin-velocity moment evolution equation is 
\begin{eqnarray}
        \frac{ D \Sigma_{ij} }{Dt} &=& 
	- \Sigma_{ij} \frac{\partial U_k}{\partial x_k}  
	- \Sigma_{ik} \frac{\partial U_j}{\partial x_k} 
	- P_{jk} \frac{\partial S_i}{\partial x_k} 	
	\nonumber \\ &&
	+ \frac{q}{m} \epsilon_{jkl} \Sigma_{ik} B_l 
	+  \frac{2 \mu}{\hbar} \epsilon_{ikl} \Sigma_{kj} B_l
	\nonumber \\ &&
	+ \mu n \frac{\partial B_i}{\partial x_j}  
	- \mu n S_i S_k \frac{\partial B_k}{\partial x_j} 
	- \frac{\partial \Lambda_{ijk}}{\partial x_k} .	
	\label{14} 
\end{eqnarray}
It is instructive to compare these fluid equations with the ones obtained in Ref.\
[\onlinecite{marklund2007}] which is based on a Madelung ansatz of the wave function. The equations 
obtained there contain additional terms which are explicitly quantum mechanical already in the
velocity and spin moment equations, for example, the Bohm potential and various spin-spin
correlation terms. In the current paper these terms are missing for two different reasons. 
Some of the terms are 'hidden' in higher order moments (see [\onlinecite{haas2005}]), and others are missing since our starting point here was the long scale length limit of the kinetic equation. 

\subsection{Two-fluid model}
The Eqs.\ \eqref{10}--\eqref{14} constitutes spin-fluid equations for a single
electron-fluid. However, since the kinetic equation is linear in  $f$, it 
is straightforward to divide the electron fluid into two parts
depending on the initial spin state, e.g.\ up- or down relative to the
magnetic field. The Eqs.\ \eqref{10}--\eqref{14} will then apply to two sets of 
electron fluids, which differ by their
initial value of the spin moment \eqref{spindens}. 
Such a division has been succesful in deriving the rate
of spin-polarization due to the spin-dependent part of the ponderomotive
force \cite{brodin2010}. In case simplified spin-fluid models without the
velocity-spin tensor are used (e.g.\ Refs.\ [\onlinecite{brodin-marklund2007}] and [\onlinecite{zamanian2009}]), such 
a division is necessary, in order to capture the basic physics of how the magnetic dipole force can
induce a spin-polarization in a plasma \cite{brodin2008}. If the model is more refined and
$\Sigma_{ij}$ is kept, on the other hand, a single fluid model will
suffice to cover this effect. Therefore the choice of a one-fluid or
two-fluid model is to a large extent a matter of taste, provided
$\Sigma_{ij}$ is kept. The advantage with the one-fluid model is that it
reduces the number of dependent variables, since only a single
electron-fluid is considered. The advantage with a two-fluid model (with
division of the electron fluid made with respect to the initial spin
state) is that much of the physics can be captured in the simpler
moments \eqref{velocity} and \eqref{spindens}, that are easier to interprete physically, rather
than in the more complicated tensor \eqref{spinvel}. It should be stressed that more highly nonlinear
and/or thermal effects may require an evolution equation for even higher order moments, i.e. for $\Lambda_{ijk}$ and higher.

\section{Linear motion perpendicular to the magnetic field}

We now consider an electron-ion plasma which is subject to an external magnetic field in the
z-direction, $\mathbf B_0 = B_0 \hat{\mathbf z}$. 
If we consider sufficiently fast dynamics the ions can be
taken as a stationary neutralizing background with number density $n_0$. To obtain a closed set of equations the closure conditions $Q_{ijk} = 0$ and $\Lambda_{ijk} = 0$ are assumed, appropriate for low temperatures. 
For perturbations of the system $n = n_0 + n_1$, $\mathbf v = \mathbf v_1$, $P_{ij} = P^{(0)}_{ij} + P^{(1)}_{ij}$, $\mathbf S = \mathbf S_0 + \mathbf S_1$, $\mathbf E = \mathbf E_1$ and $\mathbf B = \mathbf B_0 + \mathbf B_1$, with small deviance from the equilibrium (denoted by subscript 0) we may linearize the equation of motion for the electrons Eq.\ \eqref{10}--\eqref{14}. 

We will assume a thermal background distribution $f_0$. The appropriate form of this distribution function is (for sufficiently large chemical potential) a generalization of a Maxwell-Boltzmann distribution given by \cite{zamanian2010} 
\begin{equation} 
	f_0 =  \frac{n_0}{(2 \pi)^{5/2} v_T^3}  e^{- v^2/(2 v_T^2) } 
	\left[ 1 + \tanh \left( \frac{ \mu_B B_0}{k_B T} \right) \cos \theta_s \right] .
\end{equation}
Using this distribution the equilibrium pressure moment is $P_0 = n_0 k_B T$ and the equilibrium polarization density is
\begin{equation}
  \mathbf S_0 =  \tanh \left( \frac{ \mu B_0 }{k_B T} \right) \hat{ \mathbf{z}} \approx
   \frac{ \mu B_0}{k_B T} \hat{ \mathbf{z}},   
\end{equation} 
where the last approximation is valid when $\mu B_0 \ll k_B T $. Note that for electrons $\mu < 0$ so that the spins tend to align anti-parallel to the magnetic field. 

To find linear solutions we Fourier decompose with wave vector $\mathbf k = k \hat{\mathbf x}$. Looking for solutions with the electric field polarization along the $z$-axis $\mathbf E = E_z\hat{\mathbf z}$ we can in the end verify that $\sigma_{xz}$ and $\sigma_{yz}$  are zero and hence it is only necessary to keep the $\sigma_{zz}$ component of the conductivity tensor. For convenience we here use an effective conductivity tensor where the total current $j_i$, including both the free and the magnetization current, is related to the electric field by $j_i = \sigma_{ij} E_j$.  We obtain, after some algebra, the dispersion relation
\begin{eqnarray}
	\omega^2 - k^2 c^2  
	&=&  
	\omega_p^2 
	+ \frac{\omega_p^2 \hbar^2 k^2 }{2 m k_B T} \left[ \frac{\omega_c}{\omega - \omega_{cg}}
	-  \frac{\omega_c}{\omega + \omega_{cg}} \right] 
	\nonumber \\ && 
	+ \frac{\omega_p^2}{\omega_c^2} \frac{ \hbar^2 k^4}{16m^2}  
	\left[ - \frac{2 \omega_{cg} }{\omega - \omega_{cg}}
	+ \frac{ 2 \omega_{cg} }{\omega + \omega_{cg}} 
	\right. 
	\nonumber \\ &&
	+ \frac{ \omega_c + \omega_{cg}  }{\omega - \omega_{cg} - \omega_c }
	+ \frac{  \omega_{cg} - \omega_c }{\omega - \omega_{cg} + \omega_c }
	\nonumber \\ && 
	\left.
	-  \frac{ \omega_{cg} - \omega_c }{\omega + \omega_{cg} - \omega_c }
	- \frac{ \omega_c + \omega_{cg} }{\omega + \omega_{cg} + \omega_c }
	\right]
\end{eqnarray}
where we have kept only the terms that survive in the limit $T \rightarrow 0$. 
As can be seen from this there is a resonance at $\omega = \omega_{cg}$, however, we know from
the corresponding kinetic theory  \cite{zamanian2010} that we also have cyclotron modes with resonace at $\omega =
\omega_c$. 
So, unless the frequency resolution is very high it is difficult to separate this spin resonance from the classical resonance. 
Instead we consider frequencies sufficiently close to the differences $\omega \approx \Delta \omega = \omega_c (g/2 -1)$, the dispersion relation is then
\begin{equation}
	\omega^2 - k^2 c^2 = \omega_p^2 
	+ \frac{\omega_p^2}{\omega_c^2} \frac{\hbar^2 k^4}{16 m^2} 
        \frac{\Delta \omega }{\omega - \Delta \omega} .
\end{equation}
This dispersion relation is the same as obtained from the full kinetic theory in [\onlinecite{brodin2009}].
If one is interested in thermal effects it is necessary to keep the terms from the
third order mixed moment $\Lambda_{ijk}$ since thermal terms will be introduced in the second
order moment which are cancelled by a contribution from the next higher order. 
This seems to be a general complication with quantum fluid theory in the linear limit 
\cite{haas2010-1,haas2010-2}. Since the linear solutions are obtainable directly from the kinetic 
theory, perhaps the main strength of using a fluid theory is when considering non-linear motion, where 
the kinetic theory quickly becomes unmanageable.

\section{Summary and discussion}
We have in this paper derived the fluid equations for spin-1/2 particles. It is found that important
effects occur due to the correlation between spin and velocity, as covered by the velocity-spin
tensor $\Sigma_{ij}$. The closure problem of the hierarchy of fluid equations has been pointed out. 
We note that it is hard to know a priori how to close the fluid system in a given situation. For a
low-temperature linear problem, we have shown that it suffice to drop moments $\Lambda_{ijk}$ and
higher. However, the real need for a fluid model occurs in inhomogeneous and/or nonlinear problems,
when kinetic theory becomes increasingly cumbersome to apply both analytically and numerically. To
what extent dropping higher order moments is adequate in such scenarios is an open question. An
important area of future research is thus to investigate the applicability of the fluid model to
more complicated problems.

\acknowledgments
MM was supported by the Swedish Research Council Contract No.\ 2007-4422 and the European Research Council Contract No.\ 204059-QPQV.

\end{document}